\documentclass[sigconf]{acmart}
\usepackage{hyperref}
\AtBeginDocument{%
  }

\usepackage{algorithm}
\usepackage{algorithmicx}
\usepackage{algpseudocode}
\usepackage{amsmath}
\usepackage{amsfonts}

\usepackage{enumitem}
\usepackage{multirow}
\usepackage{array}
\usepackage{graphicx}
\usepackage{textcomp}
\usepackage{cleveref}
\usepackage{subcaption}
\copyrightyear{2026}
\acmYear{2026}
\setcopyright{cc}
\setcctype{by}
\acmConference[RecSys '26]{20th ACM Conference on Recommender Systems}{September 27-October 02, 2026}{Minneapolis, MN, USA}
\acmBooktitle{20th ACM Conference on Recommender Systems (RecSys '26), September 27-October 02, 2026, Minneapolis, MN, USA}
\acmDOI{10.1145/3773078.3831793}
\acmISBN{979-8-4007-2284-4/2026/09}

\begin{document}

\title{Hamiltonian Spectral-Temporal Dissipative Dynamics for Sequential Recommendation}

\author{Shuiying Liao}
\email{shuiyingl@ust.hk}
\affiliation{%
  \institution{The Hong Kong University of Science and Technology}
  \city{Clear Water Bay}
  \country{Hong Kong}
}

\author{P. Y. Mok}\authornote{Corresponding author.}
\email{tracy.mok@ust.hk}
\affiliation{%
  \institution{The Hong Kong University of Science and Technology}
  \city{Clear Water Bay}
  \country{Hong Kong}
}

\begin{abstract}
Sequential recommendation requires understanding how user preferences evolve over time, yet most existing models treat such evolution as a first order process where the next state depends solely on the current latent representation. Nevertheless, real user behavior often exhibits richer dynamics, including inertia, periodicity, and sudden shifts that cannot be fully captured by these first order assumptions. Motivated by these behavioral characteristics, we reconceptualize sequential recommendation through the lens of second order dynamical systems and introduce the \textbf{H}amiltonian \textbf{S}pectral \textbf{R}ecommender (\textbf{HSR}), which recasts preference evolution as a \emph{dissipative Hamiltonian system} in a latent phase space of position (stable preference) and momentum (short-term tendency). The linear time-invariant structure of the governing equation admits a closed-form solution in the frequency domain. 
A learnable dissipation mechanism further captures natural interest decay, while a short local impulse refinement module models abrupt behavioral fluctuations commonly observed in sparse interaction logs. 
This design jointly accounts for global periodic patterns, inertial evolution, and localized shocks, where three 
phenomena that are underrepresented in existing sequential models. Extensive experiments on three benchmark datasets demonstrate that HSR consistently outperforms state-of-the-art Transformer-based and state space model (SSM)-based recommenders. The code is available at \url{https://github.com/asaander719/HSR}.

\end{abstract}



\maketitle

\section{Introduction}
Sequential recommendation plays a central role in modern personalization systems, where the goal is to predict a user’s next interaction by understanding how their interests evolve over time. Applications such as e-commerce, short video feeds, music streaming, and online advertising rely heavily on accurate modeling of these temporal preference dynamics~\cite{liao2024hypergraph}.
Although deep learning has substantially advanced sequential recommendation, through architectures based on convolutional \cite{Caser} or recurrent \cite{gru4rec} networks, Transformers \cite{SASRec, bert4rec, diff25, MBHT, baek2025muffin}, and recently structured state space models (SSMs) \cite{Mamba4Rec,zhang2025local, zhang2025m2rec, fan2025tim4rec, patro2025mamba}, a fundamental assumption remains largely unchanged: \textit{user preference is treated as a first-order process driven solely by the current latent state.} 

\begin{figure}[t]
  \centering
  \setlength{\abovecaptionskip}{0.2cm}
  \includegraphics[width=0.95\linewidth]{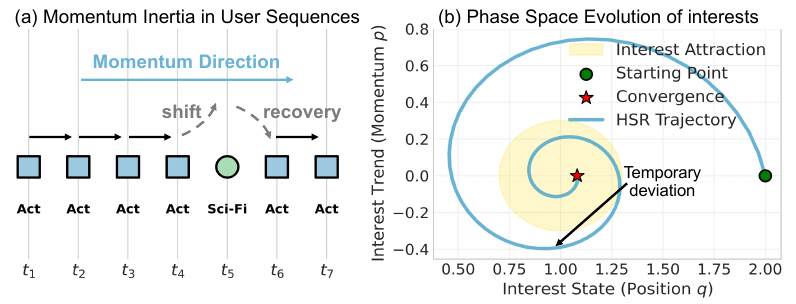}
  \caption{Main Motivation of HSR.}
  \label{fig: teaser}
\end{figure}
Nevertheless, real user behavior rarely fits this simplification. Consider the behavior pattern in Figure~\ref{fig: teaser}(a): a user watches several action movies, is briefly distracted by one science-fiction trailer, and returns to action. A first-order model would drift toward the new topic, while a human observer reads three intertwined effects: \emph{\textit{inertia}} in the action interest; a \emph{\textit{temporary deviation}}; and a \emph{\textit{damped recovery}}, which are precisely the signatures of a second-order dynamical system. Figure~\ref{fig: teaser}(b) visualizes the same behavior in phase space: the trajectory orbits an interest attractor, is knocked off course, and spirals back, a pattern no first-order recurrence can reproduce exactly.

Empirical evidence from behavioral psychology, digital consumption, and user-log analysis shows that people often display \textit{inertia} (maintaining ongoing interests over extended periods), \textit{periodicity} (seasonal or cyclical return to topics), and \textit{abrupt shifts} (sudden attention to trends or new stimuli) \cite{qu2025intent, liao2023recommendation, ding2021modeling, liao2024consistency, ding2023computational}. These diverse behaviors imply that preference evolution depends not only on the current preference state but also on how quickly that state is changing.
These observations motivate a fundamentally different modeling lens. Hamiltonian mechanics provides a mathematically principled description of second-order dynamics, with naturally interpretable constructs for inertia, oscillation, and dissipative decay~\cite{thompson2002nonlinear, arnold1989mathematical, hnn, khalil2002nonlinear, liang2025spini, okamoto2025learning}. In its dissipative form, the governing equation simultaneously captures how strongly an ongoing trend persists, how quickly transient interests fade, how firmly the trajectory is pulled toward stable preferences, and how external item exposures perturb the state. Although these concepts are foundational in physics and control, they have not been integrated into sequential recommendation.

We propose a novel \textbf{Hamiltonian Spectral Recommender (HSR)}, which recasts preference evolution as a dissipative Hamiltonian system in a latent phase space of \textit{position} (stable preference) and \textit{momentum} (short-term tendency). Three design choices make this formulation practical. (a) Since the governing equation is linear and time-invariant, it admits a closed-form solution in the frequency domain. We realize this as a learnable spectral propagator whose denominator is fixed by physics and whose numerator is data-driven, reducing propagation from an $O(T^2)$ time-domain simulation to an $O(T\log T)$ FFT. (b) A local impulse branch operating in parallel models the abrupt deviations that smooth second-order dynamics cannot explain on their own. (c) Rather than using the final position as the user representation, we perform a \emph{one-step phase-space extrapolation} that advances the state along its instantaneous velocity, which turning the predicted momentum into an actual prediction of \emph{where the user is heading}, not just \emph{where they have been}.

\begin{figure*}[t!]
  \centering
  \setlength{\abovecaptionskip}{0.2cm}
  \includegraphics[width=\linewidth]{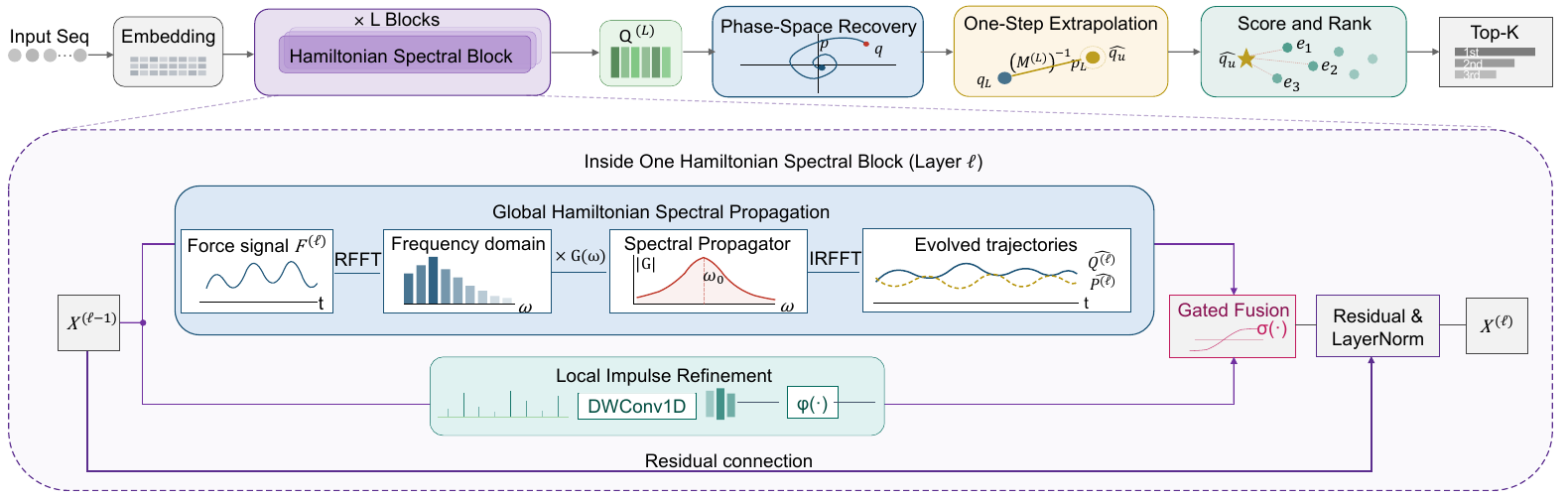}
  \caption{Overall Architecture of HSR, including the flow from input sequence through embedding, Hamiltonian spectral layer, impulse refinement, and prediction to generate personalized recommendations.}
  \label{fig: main}
\end{figure*}

The main contributions of this work are:
\begin{itemize}
    \item a new formulation of user preference evolution as a dissipative second-order dynamical system;
    \item the first \emph{Hamiltonian spectral propagator} that models long-range inertial combined with a local impulse branch that captures abrupt behavioral deviations;
    \item a \emph{one-step phase-space extrapolation} that uses the recovered momentum as the native mechanism for next-step prediction;
    \item extensive empirical evidence showing that this perspective provides meaningful gains over existing first-order approaches.
\end{itemize}

\section{Related Work}
Existing sequential recommendation methods can be broadly categorized into three research directions: neural sequence modeling, frequency-domain representation learning, and structured state-space modeling. Since HSR draws inspiration from spectral analysis and dynamical systems, we briefly review these closely related lines of work below.

\textbf{Neural Models for Sequential Recommendation} Early deep-learning approaches to sequential recommendation focused on learning temporal dependencies using convolutional and recurrent architectures. Models such as Caser \cite{Caser} and GRU4Rec \cite{Hidasi2015SessionbasedRW} demonstrated that CNNs and gated recurrent units can effectively capture local or auto-regressive behavioral patterns. However, these architectures often struggle with long-range dependencies due to vanishing gradients and limited receptive fields.
The introduction of self-attention mechanisms marked a significant milestone. SASRec \cite{SASRec}, BERT4Rec \cite{bert4rec} and extensions \cite{diff25} demonstrated that Transformer-based architectures can effectively capture global dependencies among user interactions. To improve scalability on long interaction histories, recent architectures such as HSTU \cite{hstu} replace quadratic self-attention with linear-complexity projection mechanisms, enabling efficient long-context modeling while maintaining strong recommendation performance. 
\textit{Although these neural architectures have achieved remarkable success, they primarily model preference evolution through latent state transformations and do not explicitly characterize higher-order behavioral dynamics such as inertia, oscillation, or momentum-driven preference shifts.}

\textbf{Frequency-Domain and Spectral Approaches} Another line of research explores the use of spectral representations to capture periodicity and denoise interaction sequences. Recent examples include methods \cite{diff25, baek2025muffin, he2026exploiting, xu2026wavelet, dang2026pay} that operate in the frequency domain to filter out short-term fluctuations or amplify stable patterns. 
\textit{Although these methods demonstrate that frequency-domain information can benefit sequential recommendation, spectral transformations are typically used as feature enhancement or filtering mechanisms. As a result, the frequency components themselves often lack an explicit dynamical interpretation of user-interest evolution.}

\textbf{Structured State-Space Models (SSMs)} represent a more recent trend toward scalable sequence modeling. By parameterizing linear time-invariant (LTI) systems, models such as Mamba4Rec \cite{Mamba4Rec} and its variants \cite{SIGMA, qu2024ssd4rec}  efficiently process long sequences with linear-time complexity. 
\textit{However, existing SSMs primarily rely on first-order state transitions. Although effective for modeling temporal dependencies, they do not explicitly distinguish between preference state and preference velocity, making it difficult to directly represent behavioral inertia or perform phase-space extrapolation at prediction time.}

\section{Method}
\label{sec:method}

\subsection{Problem Formulation}
Let $\mathcal{U}$ denote the user set and $\mathcal{V}$ the item set. A user $u \in \mathcal{U}$, is associated with a historical interaction sequence
    $S_u = [v_1, v_2, \dots, v_T],$ 
where $v_t \in \mathcal{V}$ is the item interacted at time $t$, and $T$ is the length of the sequence. Each item $v$ corresponds to a learnable embedding $e_v \in \mathbb{R}^d$. 
Mapping $S_u$ through this embedding table yields \begin{equation}
    X = [x_1, \dots, x_T]^\top \in \mathbb{R}^{T \times d},
    \quad x_t = e_{v_t}.
    \label{eq:input-emb}
\end{equation}
The goal of sequential recommendation is to predict the next item $v_{T+1}$ that the user is most likely to interact with, given the observed prefix $S_u$.

\subsection{Overview}
Figure~\ref{fig: main} illustrates the main architecture of HSR. At a high level, HSR models the temporal evolution of user preference as a \emph{dissipative Hamiltonian dynamical system} in a latent space. The architecture consists of five components:
(i) \textbf{Embedding layer} maps discrete item IDs to dense vectors (eq.~\eqref{eq:input-emb}) and injects temporal order via learnable position encodings;
(ii) \textbf{Hamiltonian Spectral Encoder}: a stack of $L$ identical Hamiltonian blocks, each performing both global spectral evolution in the frequency domain and local impulse modeling in the time domain, with gated fusion and residual normalization;
(iii) \textbf{Final Phase-Space Recovery} re-derives the terminal momentum from the final fused position trajectory through an RFFT--derivative--IRFFT pass, ensuring that position and momentum used for prediction correspond to the \textit{same} final latent state;
(iv) \textbf{One-Step Extrapolation} advances the terminal state by a single Euler step along its instantaneous momentum, yielding the predicted next-step latent preference;
(v) \textbf{Prediction Layer} scores each candidate item by the compatibility between  and the item embedding, and ranks all items for top-$K$ recommendation.

\subsection{Primer}
\subsubsection{Hamiltonian Dissipative Dynamics in Phase-Space}
To capture the complex evolution of user interests—such as inertia and natural decay—we depart from first-order transitions. We represent the user's latent state in a conjugated phase-space $(q, p)$, where $q(t) \in \mathbb{R}^d$ denotes the \textit{position} (stable preference) and $p(t) \in \mathbb{R}^d$ denotes the \textit{momentum} (short-term behavioral tendency).

The evolution of the user system is governed by a dissipative Hamiltonian system. For each latent dimension $i$, the dynamics are described by:
\begin{equation}
    \dot{q}_i(t) = p_i(t)/{m_i}, \quad \dot{p}_i(t) = -\kappa_i q_i(t) - c_i \dot{q}_i(t) + f_i(t)
    \label{eq:hamiltonian_dynamics}
\end{equation}
where $m_i, c_i, \kappa_i$ are learnable parameters representing \textit{mass} (inertia), \textit{damping} (forgetting rate), and \textit{stiffness} (preference stability), respectively. $f_i(t)$ represents the external force exerted by item interactions. 
By taking the time derivative of $p_i = m_i \dot{q}_i$ gives $\dot{p}_i = m_i \ddot{q}_i$, and substituting into eq.~\eqref{eq:hamiltonian_dynamics}, we obtain the second-order damped driven oscillator equation:
\begin{equation}
    m_i \ddot{q}_i(t) + c_i \dot{q}_i(t) + \kappa_i q_i(t) = f_i(t)
    \label{eq:second_order}
\end{equation}

\subsubsection{Spectral Propagator and Hamiltonian Filtering}
Solving eq.~\eqref{eq:second_order} in the time domain for long sequences is computationally expensive. However, by applying the Fourier Transform $\mathcal{F}[\cdot]$, the differential equation transforms into an algebraic relation in the \textit{frequency domain ($\omega$)}:
\begin{equation}
    (-m_i \omega^2 + i c_i \omega + \kappa_i) \hat{q}_i(\omega) = \hat{f}_i(\omega),
\end{equation}
which allows us to define the \textit{\textbf{Spectral Propagator}} $G_i(\omega)$, which acts as a complex filter that shapes the user's preference trajectory:
\begin{equation}
    G_i(\omega) = \hat{q}_i(\omega)/{\hat{f}_i(\omega)} = 1/({\kappa_i - m_i \omega^2 + i c_i \omega}).
    \label{eq:spectral_filter}
\end{equation}

To compute the final state for prediction, we reconstruct the momentum $p_i(t)$ to preserve the Hamiltonian structure. In the spectral domain, the momentum is derived as $\hat{p}_i(\omega) = m_i (i\omega) \hat{q}_i(\omega)$. 

\subsection{HSR Layer Implementation}
\label{{sec:hsr-block}}

Building upon the spectral propagator in eq.~\eqref{eq:spectral_filter}, we implement the HSR layer using a Discrete Fourier Transform (DFT). 
A block takes 
$
    X^{(\ell-1)}\in \mathbb{R}^{T \times d}
$
as input, and outputs an updated sequence
$
    X^{(\ell)}\in \mathbb{R}^{T \times d}.
$
Each block contains three stages:
(I) \emph{projection mapping} from the latent sequence to a driving force and stochastic excitation;
(II) \emph{global Hamiltonian spectral propagation} according to a Hamiltonian propagator; and
(III) \emph{local impulse refinement with gated fusion} in the time domain.

\paragraph{\textbf{Stage I: Projection Mapping.}}
The input sequence is projected into three branches: a driving-force branch, a local-impulse refinement, and a gate branch. We first compute:
\begin{equation}
    \left[\boldsymbol{F}^{(\ell)}_{in}, \boldsymbol{U}^{(\ell)}_{in}, \boldsymbol{A}^{(\ell)} \right] = X^{(\ell-1)} \mathbf{W}^{(\ell)}_{in} + \mathbf{b}^{(\ell)}_{in},
    \label{eq:projection_params}
\end{equation}
where 
$
    \mathbf{W}^{(\ell)}_{in} \in \mathbb{R}^{d \times 3d},
    \qquad
    \mathbf{b}^{(\ell)}_{in} \in \mathbb{R}^{3d}.
$
We then split the projected tensor along the channel dimension and define:
\begin{equation}
    \boldsymbol{F}^{(\ell)} = \phi_f \left(\boldsymbol{F}^{(\ell)}_{in} \right),
    \quad
    \boldsymbol{U}^{(\ell)} = \phi_u \left(\boldsymbol{U}^{(\ell)}_{in} \right),
    \quad
    \boldsymbol{g}^{(\ell)} =\sigma \left(\boldsymbol{A}^{(\ell)} \right),
\end{equation}
where $\phi_f(\cdot)$ and $\phi_u(\cdot)$ are activation function such as GELU, and $\sigma(\cdot)$ is the logistic sigmoid. The interpretation is as follows.
$\boldsymbol{F}^{(\ell)}$ is the global \emph{driving force} signal, entering the second-order dynamics.
$\boldsymbol{U}^{(\ell)}$ models local \emph{short-term impulse} signal, such as bursty clicks or transient curiosity.
$\boldsymbol{g}^{(\ell)}$ is a \emph{gating signal} adaptively controls how strongly the fused signal is expressed in the block output. Importantly, we do not directly project the input into a free momentum channel. Instead, momentum is recovered from the evolved position trajectory through second-order dynamics (later eq.~\eqref{eq:phat_discrete}), which preserves phase-space consistency.

\paragraph{\textbf{Stage II: Global Hamiltonian Spectral Propagation.}}
For each latent dimension, we approximate eq.~\eqref{eq:second_order} by a discrete linear time-invariant system whose solution is represented in the frequency domain.
Specifically, we apply the real-valued fast Fourier transform (RFFT) to the driving-force sequence $\widehat{\boldsymbol{F}}^{(\ell)} \in  \mathbb{R}^{T \times d}$ along the temporal dimension. Here, $T$ is the sequence length and $d$ is the hidden dimension. For each latent channel $i \in \{1,\dots,d\}$, the vector $F^{(\ell)}_{:,i} \in \mathbb{R}^{T}$ is treated as a one-dimensional discrete-time signal over the user history. The temporal Fourier transform converts this signal from the time domain into the frequency domain, which allows us to model long-range and periodic preference dynamics through a frequency-dependent propagator.
Formally, let $N_f$ denotes the number of spectral coefficients, 
$
    \widehat{\boldsymbol{F}}^{(\ell)} = \mathcal{F}_{\mathrm{RFFT}}(\boldsymbol{F}^{(\ell)})
    \in \mathbb{C}^{d \times N_f},
    N_f = \lfloor T/2 \rfloor + 1.
$
Here $ \mathcal{F}_{\mathrm{RFFT}}(\cdot)$ denotes the real FFT applied along the temporal axis. Since the input signal is real-valued, its discrete Fourier spectrum satisfies Hermitian symmetry, and therefore only the non-redundant non-negative frequency modes need to be retained.

Let $n= 0,1, \dots, N_f-1$ denote the discrete frequency index, then the corresponding angular frequency is defined as
$
    \omega_n = 2\pi n/T,
$
where we assume a unit sampling between adjacent interaction steps. Each spectral coefficient $\widehat{\boldsymbol{F}}^{(\ell)}_{n,i} \in \mathbb{C}$ is a complex number associated with frequency mode $n$ and latent channel $i$. Its magnitude reflects how strongly the input sequence excites that frequency component, while its phase encodes temporal alignment information.
For each layer $\ell$ and latent dimension $i$, we parameterize the physical coefficients by unconstrained variables $\bar{m}^{(\ell)}_i$, $\bar{c}^{(\ell)}_i$, and $\bar{\kappa}^{(\ell)}_i$,
and enforce positivity through:
\begin{equation}
m_i^{(\ell)} = softplus\!\left(\bar{m}^{(\ell)}_i\right) + \varepsilon_m,
\end{equation}
\begin{equation}
c_i^{(\ell)} = softplus\!\left(\bar{c}^{(\ell)}_i\right),
\quad
\kappa_i^{(\ell)} = softplus\!\left(\bar{\kappa}^{(\ell)}_i\right),
\label{eq:physical_params}
\end{equation}
where $softplus(\boldsymbol{x})=log(1+e^{\boldsymbol{x}})$, which is a smooth positive-valued mapping function. This parameterization ensures that the learned physical coefficients remain interpretable and numerically stable, as mass is strictly positive, while damping and stiffness are non-negative.
Based on the frequency-domain form of the damped second-order system, we define the spectral response term:
\begin{equation}
D_{n,i}^{(\ell)} =
\kappa_i^{(\ell)} -
m_i^{(\ell)} \omega_n^2 +
\mathrm{i} c_i^{(\ell)} \omega_n + \varepsilon,
\label{eq:denominator}
\end{equation}
where $\varepsilon > 0$ is a small constant for numerical stability.
The term $D_{n,i}^{(\ell)}$ can be interpreted as the frequency-domain impedance of the $i$-th latent channel at the $n$-th Fourier mode.
$\kappa_i^{(\ell)}$ contributes the restoring effect,
$m_i^{(\ell)} \omega_n^2$ captures inertial resistance to fast oscillation,
and $\mathrm{i} c_i^{(\ell)} \omega_n$ introduces damping and phase lag.

To allow additional flexibility beyond the idealized physical model, we further introduce a learnable complex numerator $\Psi_{n,i}^{(\ell)} \in \mathbb{C}$, and define the discrete Hamiltonian-inspired propagator as:
\begin{equation}
\mathcal{G}_{n,i}^{(\ell)} =
\Psi_{n,i}^{(\ell)}/{D_{n,i}^{(\ell)}}.
\label{eq:discrete_propagator}
\end{equation}
Here, $\mathcal{G}_{n,i}^{(\ell)}$ specifies how the $n$-th frequency component of the driving-force branch is transformed in the $i$-th latent channel.
The evolved position spectrum is then computed by:
\begin{equation}
\widehat{Q}_{n,i}^{(\ell)} =
\mathcal{G}_{n,i}^{(\ell)} \widehat{F}_{n,i}^{(\ell)}.
\label{eq:qhat_discrete}
\end{equation}
This equation is the discrete counterpart of the continuous spectral solution
of eq.~\eqref{eq:spectral_filter},
the propagator acts as a frequency-dependent filter that selectively amplifies, attenuates, and phase-shifts different temporal patterns in the input sequence.

Next, instead of learning momentum as an independent feature stream, we recover it from the same evolved position spectrum through the derivative relation implied by the second-order dynamics:
\begin{equation}
\widehat{P}_{n,i}^{(\ell)} =
m_i^{(\ell)} (\mathrm{i}\omega_n) \widehat{Q}_{n,i}^{(\ell)}.
\label{eq:phat_discrete}
\end{equation}
This step ensures that position and momentum remain coupled through a physically meaningful relation rather than becoming two unrelated latent representations.
We then reconstruct the time-domain trajectories by inverse FFT:
\begin{equation}
\widetilde{\boldsymbol{Q}}^{(\ell)} =
\mathcal{F}_{\mathrm{RFFT}}^{-1}\!\left(\widehat{\boldsymbol{Q}}^{(\ell)}\right) \in \mathbb{R}^{d \times T},
\quad
\widetilde{\boldsymbol{P}}^{(\ell)} =
\mathcal{F}_{\mathrm{RFFT}}^{-1}\!\left(\widehat{\boldsymbol{P}}^{(\ell)}\right) \in \mathbb{R}^{d \times T}.
\label{eq:qp_tilde}
\end{equation}
Here, $\widetilde{\boldsymbol{Q}}^{(\ell)}$ is the globally evolved position trajectory and $\widehat{\boldsymbol{P}}^{(\ell)}$ is the corresponding momentum trajectory induced by the same spectral dynamics.

\paragraph{\textbf{Stage III: Local Impulse Refinement and Gated Fusion.}}

The global spectral branch captures smooth, long-range preference evolution. However, recommendation logs also contain abrupt local deviations that are difficult to explain using only a globally smooth trajectory. To model such effects, we refine the representation with the local impulse refinement.

We first apply a depthwise temporal convolution to the local impulse refinement:
\begin{equation}
\boldsymbol{H}_{\mathrm{loc}}^{(\ell)} =
DWConv1D\left(\boldsymbol{U}^{(\ell)}\right) \in \mathbb{R}^{T \times d}.
\label{eq:local_impulse}
\end{equation}
We then fuse the global position trajectory with the local refinement:
\begin{equation}
\boldsymbol{Z}^{(\ell)} =
\left( \widetilde{\boldsymbol{Q}}^{(\ell)} +
\phi\left(\boldsymbol{H}_{\mathrm{loc}}^{(\ell)}\right) \right)
\odot \boldsymbol{g}^{(\ell)},
\label{eq:fusion}
\end{equation}
where $\phi(\cdot)$ is a smooth nonlinearity such as GELU, and $\odot$ denotes element-wise multiplication. Then the block output is given by:
\begin{equation}
\boldsymbol{X}^{(\ell)} = LayerNorm \left(
\boldsymbol{X}^{(\ell-1)} +
Dropout\!\left(\boldsymbol{Z}^{(\ell)} \boldsymbol{W}_o^{(\ell)}\right) \right),
\label{eq:block_update}
\end{equation}
where $\boldsymbol{W}_o^{(\ell)} \in \mathbb{R}^{d \times d}$.
Eq.~\eqref{eq:block_update} preserves the original architectural intuition while making the roles of different branches explicit, as the spectral branch models smooth global evolution,
the local impulse refinement models short-term local shocks,
and the gate branch modulates the final fused representation.

\subsection{Stacking Blocks and Final Phase-Space Recovery}
We stack $L$ Hamiltonian spectral blocks
$
[\boldsymbol{X}^{(0)} \rightarrow \boldsymbol{X}^{(1)} \rightarrow \cdots \rightarrow \boldsymbol{X}^{(L)}].
$
After local impulse refinement, gating, residual fusion, and normalization, the final hidden sequence $\boldsymbol{X}^{(L)}$ is no longer identical to the raw spectral position $\widetilde{\boldsymbol{Q}}^{(L)}$ from the last block.
Therefore, if one uses the final position from $\boldsymbol{X}^{(L)}$ but the momentum from an earlier spectral branch, the resulting pair would not correspond to the same final state. We use a dedicated terminal mass $\boldsymbol{m}$ trained jointly with the prediction head to recover momentum from $\boldsymbol{X}^{(L)}$. To ensure phase-space consistency at the recommendation head, we explicitly define the final fused position trajectory as
$
\boldsymbol{Q}^{(L)} := \boldsymbol{X}^{(L)}
$, $\widehat{\boldsymbol{Q}}^{(L)} =
\mathcal{F}_{\mathrm{RFFT}}\left(\boldsymbol{Q}^{(L)}\right)$.
We then recompute the final momentum from this same fused trajectory in the frequency domain as:
\begin{equation}
\widehat{P}_{n,i}^{(L)} =
{m}_i^{(L)} (\mathrm{i}\omega_n)\widehat{Q}_{n,i}^{(L)},
\quad
\boldsymbol{P}^{(L)} =
\mathcal{F}_{\mathrm{RFFT}}^{-1}\left(\widehat{\boldsymbol{P}}^{(L)}\right).
\label{eq:final_p}
\end{equation}
Finally, we define the terminal phase-space state as:
\begin{equation}
\boldsymbol{q}_L = \boldsymbol{Q}_T^{(L)} \in \mathbb{R}^d,
\qquad
\boldsymbol{p}_L = \boldsymbol{P}_T^{(L)} \in \mathbb{R}^d.
\label{eq:terminal_state}
\end{equation}
This step allows the model to keep the recommendation specific local impulse refinement, while still ensuring that the position and momentum used for final prediction are derived from the same final latent trajectory.


\subsection{Phase-Space Extrapolation and Prediction}
\label{subsec:prediction}


The stacked Hamiltonian spectral blocks recover a terminal latent phase-space state that summarizes both the user's current preference configuration, and its instantaneous evolution trend. As sequential recommendation is a next-step prediction problem rather than a final-state reconstruction problem. We therefore perform a one-step forward extrapolation in phase space to estimate where the user preference is likely to move next. The terminal position captures the current interest location, while the terminal momentum captures the direction and strength of short-term preference change. By analyzing momentum and advancing the state by one discrete time step, we obtain a predicted next-step latent preference for candidate ranking.

To predict the user's interest at the next time step, we perform a one-step Euler extrapolation using the learned mass parameters. 
We collect mass parameter $m_i$ parameters into a diagonal mass matrix
$
\boldsymbol{M} = \mathrm{diag}(\boldsymbol{m}_1,\dots,\boldsymbol{m}_d),
$
and denotes its inverse by $\boldsymbol{M}^{-1}$. 
We predict the next-step latent preference state by one-step phase-space extrapolation, where $\Delta t > 0$ is the step size, $
\boldsymbol{M}^{(L)} = \mathrm{diag}(\boldsymbol{m}_1^{(L)},\dots,\boldsymbol{m}_d^{(L)})
$ is the mass matrix inherited from the L-th block:
\begin{equation}
\widehat{\boldsymbol{q}}_u =
\boldsymbol{q}_L + \Delta t \left(\boldsymbol{M}^{(L)}\right)^{-1} \boldsymbol{p}_L,
\label{eq:extrapolation}
\end{equation}




\paragraph{Training Loss.}
As $\mathbf{\hat{q}}_u$ be the extrapolated sequence representation for user $u$ and $\mathbf{e}_v \in \mathbb{R}^d$ be the embedding of item $v \in \mathcal{V}$. We compute the prediction probability via a softmax over dot-product scores. The entire model is optimized end-to-end by minimizing the following function for the ground-truth next item $v^+$:
\begin{equation}
    \mathcal{L} = - \sum_{(u, v^+)\in\mathcal{D}} \log \frac{\exp(\mathbf{\hat{q}}_u^\top \mathbf{e}_{v^+})}{\sum_{v' \in \mathcal{V}} \exp(\mathbf{\hat{q}}_u^\top \mathbf{e}_{v'})} + \lambda \|\Theta\|_2^2,
    \label{eq:hsr-ce}
\end{equation}
where $\mathcal{D}$ is the set of user-item interaction pairs, and $\lambda$ is the weight-decay coefficient.
All components, from the Hamiltonian spectral blocks to the phase, and space scoring, are optimized end-to-end.

\section{Experiments}
\begin{table}[t]
\centering
\setlength{\tabcolsep}{0.1mm}
\caption{Dataset statistics.}
\label{tab:dataset_stats}
\begin{tabular}{lcccc}
\toprule
\textbf{Dataset} & \textbf{\# Users} & \textbf{\# Items} & \textbf{\# Interactions} & \textbf{Sparsity} \\
\midrule
MovieLens-1M \cite{harper2015movielens}& 6,041 & 3,417 & 999,611  & 95.53\%\\
Amazon-Beauty \cite{mcauley2015image} & 22,364 & 12,102 & 198,502  & 99.93\%\\
Amazon-Video-Games \cite{mcauley2015image} & 24,304 & 10,673 & 231,780 & 99.91\%\\
\bottomrule
\end{tabular}
\end{table}

\begin{table*}[t!]
\centering
\setlength{\tabcolsep}{1mm}
\caption{Overall Performance Comparison of Different Methods. \textbf{Bold} data indicate the best results, and \underline{\textit{underlined}} ones are the second best results.}
\label{tab:performance}
\begin{tabular}{lccccccccc}
\toprule
\multirow{2}{*}{\textbf{Method}} & \multicolumn{3}{c|}{\textbf{Amazon-Beauty}} & \multicolumn{3}{c|}{\textbf{Amazon-Video-Games}} & \multicolumn{3}{c}{\textbf{MovieLens-1M}} \\
\cmidrule(lr){2-4} \cmidrule(lr){5-7} \cmidrule(lr){8-10}
 & \textbf{Hit@10} & \textbf{NDCG@10} & \textbf{MRR@10} & \textbf{Hit@10} & \textbf{NDCG@10} & \textbf{MRR@10} & \textbf{Hit@10} & \textbf{NDCG@10} & \textbf{MRR@10}\\
\midrule
\textbf{Caser} \cite{Caser}~\textsuperscript{(WSDM'18)} & 0.0531 & 0.0294 & 0.0222 & 0.0891 & 0.0460 & 0.0330 & 0.2892 & 0.1714 & 0.1354 \\
\midrule
\textbf{GRU4Rec} \cite{Hidasi2015SessionbasedRW}~\textsuperscript{(ICLR'16)} & 0.0606 & 0.0332 & 0.0249 & 0.1030 & 0.0536 & 0.0380 & 0.2934 & 0.1642 & 0.1249 \\
\midrule
\textbf{NARM} \cite{li2017neural}~\textsuperscript{(CIKM'17)} & 0.0627 & 0.0347 & 0.0262 & 0.1032 & 0.0530 & 0.0379 & 0.2735 & 0.1506 & 0.1132\\
\textbf{SASRec} \cite{SASRec}~\textsuperscript{(ICDM'18)} & 0.0847 & 0.0425 & 0.0296 & 0.1168 & 0.0571 & 0.0390 & 0.2977 & 0.1687 & 0.1294\\
\textbf{BERT4Rec} \cite{bert4rec}~\textsuperscript{(CIKM'19)} & 0.0760 & 0.0393 & 0.0282 & 0.1053 & 0.0538 & 0.0381 & 0.3098 & 0.1764 & 0.1357 \\
\textbf{SG-MST} \cite{liao2024hypergraph}~\textsuperscript{(ICDM'24)} & 0.0819 & 0.0388 & 0.0285 &0.1182&0.0589& 0.0410 &0.3171&0.1828&0.1417\\
\textbf{HSTU} \cite{hstu}~\textsuperscript{(Meta'24)} &0.0879 &0.0419 &0.0280 &0.1317& 0.0624&0.0403
&\textbf{0.3268}&\underline{\textit{0.1863}}&\underline{\textit{0.1434}}\\
\textbf{DIFF} \cite{diff25}~\textsuperscript{(SIGIR'25)} & \underline{\textit{0.0935}} 
& \underline{\textit{0.0527}} 
& \underline{\textit{0.0406}} &\underline{\textit{0.1319}} &\underline{\textit{0.0681}} &\underline{\textit{0.0488}} &0.3126 &0.1746 &0.1327 \\
\midrule
\textbf{Mamba4Rec} \cite{Mamba4Rec}~\textsuperscript{(KDD'24)} & 0.0812 & 0.0461 & 0.0362& 0.1152 & 0.0603 & 0.0438 &  0.3207 & 0.1836 & 0.1417 \\
\textbf{SIGMA} \cite{SIGMA}~\textsuperscript{(AAAI'25)} & 0.0678 & 0.0376 & 0.0284 & 0.1239& 0.0592 & 0.0430 & 0.3136 & 0.1777 & 0.1361 \\
\textbf{SSD4Rec} \cite{qu2024ssd4rec}~\textsuperscript{(TIS'25)} & 0.0806 & 0.0423 & 0.0306 & 0.1099 & 0.0611 & 0.0464 & 0.3199 & 0.1841 & 0.1425 \\
\midrule
\textbf{HSR} (Ours) & \textbf{0.0952} & \textbf{0.0566} & \textbf{0.0448} & \textbf{0.1348} & \textbf{0.0753} & \textbf{0.0573} & \underline{\textit{0.3255}} & \textbf{0.1907} & \textbf{0.1495} \\
Gain & 1.82\% & 7.40\% & 10.34\% & 2.20\% & 10.57\% & 17.42\% & -0.39\% & 2.36\% & 4.25\% \\
\bottomrule
\end{tabular}
\label{tab:overall}
\end{table*}
\subsection{Experimental Settings}

\subsubsection{Datasets and Preprocessing}
We evaluate HSR on three widely adopted sequential recommendation datasets (Table~\ref{tab:dataset_stats}), each presenting distinct modeling challenges: 
\begin{itemize}
    
\item \textbf{Amazon-Beauty} and \textbf{Amazon-Video-Games} \cite{mcauley2015image}: Two extremely sparse datasets characterized by short, noisy interaction histories and abrupt behavioral shifts. These datasets naturally reflect irregular temporal patterns and provide strong motivation for modeling inertia, damping, and impulse-driven corrections. 
\item \textbf{MovieLens-1M} \cite{harper2015movielens}: A dense dataset with relatively frequent user interactions. Sequences are moderately long and stable, providing a testbed for understanding how HSR behaves under well-sampled preference trajectories.
\end{itemize}

For all datasets, we adopt the widely used leave-one-out evaluation setup. The final interaction of each user forms the test instance, while the penultimate interaction is used for validation. Following prior work \cite{bert4rec, Mamba4Rec, liao2024hypergraph}, we set the maximum sequence length to 200 for MovieLens and 50 for the Amazon datasets to balance sequence coverage and computational practicality.

\subsubsection{Implementation}
All models are implemented in PyTorch to ensure  reproducibility and fair comparison. We evaluate a diverse set of recommendation architectures, including convolutional-, recurrent-, Transformer-, and Mamba-based models. All models are optimized with Adam~\cite{Adam}. Mamba-based models follow recommended hyperparameters, including a state dimension of 32 and a local convolution width of 4. Transformer-based models also record their best performance based on their original report or recommended settings. HSR employs $L$ Hamiltonian spectral blocks, each operating via real FFTs over the temporal axis. Spectral parameters $(m_i, c_i,\kappa_i)$ are learnable and optimized jointly with all other model components. A batch size of 2048 is used for training, and a dropout rate of 0.2 is applied within each block for regularization. 

\subsection{Overall Performance}
\label{sec:overall}
HSR achieves the best performance across most of metrics on three datasets, as reported in Table~\ref{tab:overall}. The improvements are most notable on the Amazon-Beauty and Amazon-Video-Games datasets that are characterized by extreme sparsity and irregular behavior. This gain is particularly pronounced despite HSR using significantly fewer parameters. The Hamiltonian spectral formulation, with its explicit modeling of inertia and periodic frequency behavior of HSR provides a more suitable inductive bias for both sparse and dense environments. 

\paragraph{Comparison with attention\-based methods.}
Among attention-based models, DIFF \cite{diff25} and HSTU \cite{hstu} are the strongest baselines. 
Although attention-based models excel at pattern matching when item-level signals are clean, their quadratic cost and position-wise normalization make them less robust. HSR attains the best performance on all metrics and all datasets. The relatively larger improvement in NDCG and MRR suggests that HSR is particularly effective at re-ranking the top of the list, which is consistent with its design as a dynamics-aware model that emphasizes the next-step trajectory rather than only matching static patterns. 


\begin{figure*}[t]
  \centering
  \setlength{\abovecaptionskip}{0.2cm}
  \includegraphics[width=0.95\linewidth]{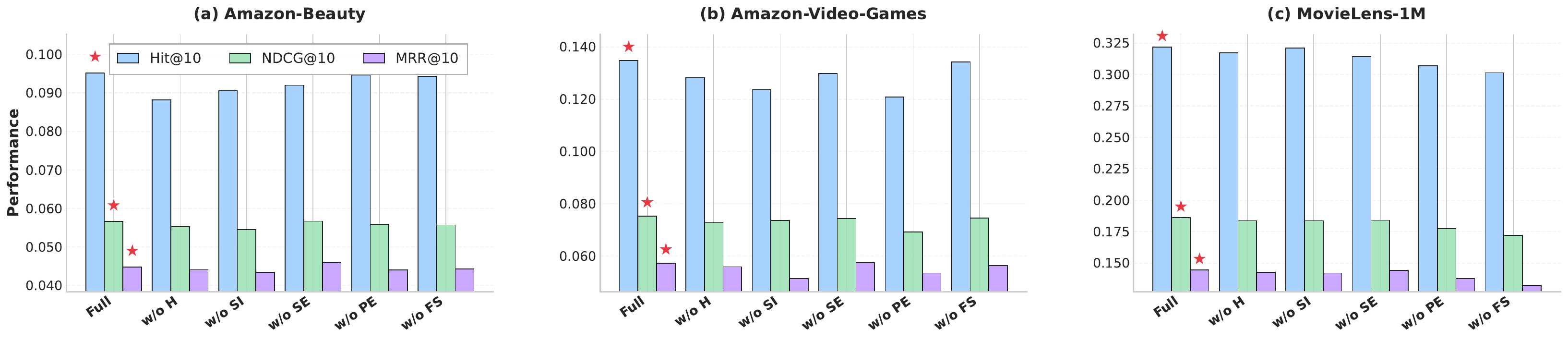}
  \caption{Model Variant Comparison, red stars indicate the best-performing variant on each metric.}
  \label{fig: ablation}
\end{figure*}

\paragraph{Comparison with SSM-based methods.}
On the dense MovieLens-1M dataset, where long-term preference structures are more stable and extensive historical context is available, SSM-based methods (Mamba4Rec \cite{Mamba4Rec}, SIGMA \cite{SIGMA}, and SSD4Rec \cite{qu2024ssd4rec}) appear outperform attention-based or transformer-based models. Existing SSM-based models with first-order LTI dynamics, and propagate information with a stable state update that can respond very quickly over long horizons. Nevertheless, their performance highly depend on the initial hidden state when sequences are short, and more sensitive to local noise. Comparatively, HSR remains competitive and surpasses all baselines. Even with fewer parameters than deep Transformer stacks, HSR’s dynamical propagation yields competitive or superior performance, especially on NDCG and MRR, which reflect ranking sensitivity. This demonstrates that second-order modeling is not merely a remedy for sparsity but provides a robust general-purpose framework for sequential preference prediction. 

\subsection{Component-wise Ablation Analysis}

\label{sec:ablation}

We perform a set of ablations to examine the contribution of each HSR component: 
\begin{itemize}
    \item \textbf{w/o Hamiltonian (H):} Replacing the entire Hamiltonian spectral block with a residual feedforward layer equipped with a depthwise temporal convolution of the same kernel size. To ensure a fair parameter budget, we set the FFN hidden dimension to $d$ and remove the gating branch. This variant explicitly removes the second-order phase-space dynamics.
    
    \item \textbf{w/o Short-Impulse (SI):} Removing the local impulse refinement module. The $\boldsymbol{U}^{(\ell)}$ branch output is simply discarded, modifying the gated fusion in eq.~\eqref{eq:fusion} to $\boldsymbol{Z}^{(\ell)} =\widetilde{\boldsymbol{Q}}^{(\ell)} \odot \boldsymbol{g}^{(\ell)}$.
    
    \item \textbf{w/o Spectral Evolution (SE):} Bypassing the spectral propagator by  $\widetilde{\boldsymbol{Q}}^{(\ell)} \approx \boldsymbol{F}^{(\ell)}$ in eq.~\eqref{eq:qhat_discrete}, disabling frequency- depen\-dent filtering while preserving the overall block topology.
    
    \item \textbf{w/o Phase-Space Extrapolation (PE):} Disabling the mom\-entum-driven prediction in eq.~\eqref{eq:extrapolation}. We use only the terminal position $\boldsymbol{q}_L$ and ignore the momentum $\boldsymbol{p}_L$, setting $\hat{\boldsymbol{q}}_u = \boldsymbol{q}_L$.
    
    \item \textbf{w/ Fourier Spectral (FS):} Replacing the learnable complex numerator with the purely Fourier-based Green's function. We fix $\boldsymbol{\Psi}_{n,i}^{(\ell)}$ to 1 in eq.~\eqref{eq:discrete_propagator}, so that $\boldsymbol{\mathcal{G}}_{n,i}^{(\ell)}= 1/\boldsymbol{D}_{n,i}^{(\ell)}$ depends solely on the physically motivated denominator. 
\end{itemize}
   
\Cref{fig: ablation} summarizes the ablations study results. These ablations highlight that each element: Hamiltonian structure; impulse refinement; spectral propagation; extrapolated prediction; and flexible spectral response, plays a distinct and meaningful role in the final performance of HSR. 

\begin{figure*}[t]
  \centering
  \setlength{\abovecaptionskip}{0.2cm}
  \includegraphics[width=0.95\linewidth]{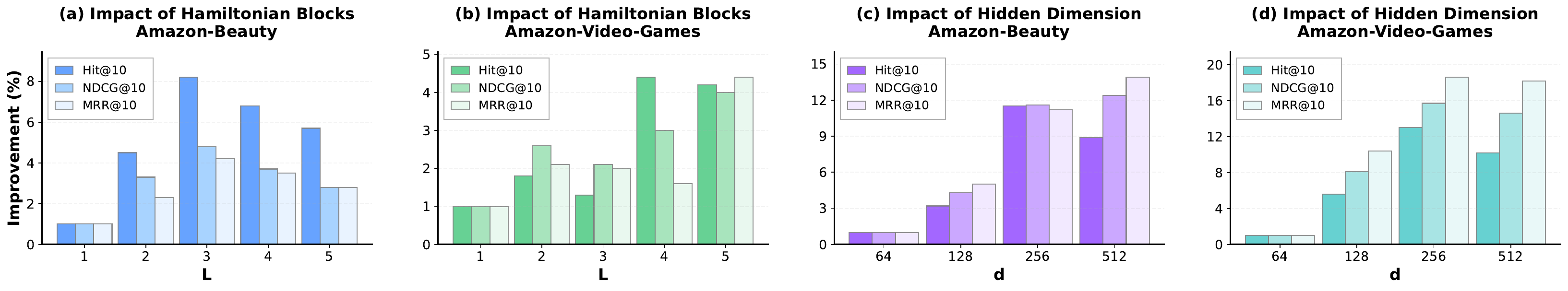}
  \caption{Hyperparameter Sensitivity Analysis: Improvement \% computed relative to $L=1$(resp. $d=64$).}
  \label{fig: hyper}
\end{figure*}

HSR-Full achieves the best performance, and removing any single module leads to a degradation in at least one metric. The largest drops appear when taking away the Hamiltonian structure or the local impulse refinement. Specifically, \textit{HSR w/o Hamiltonian} consistently underperforms HSR-Full, confirming that merely stacking residual convolutions cannot replace second-order Hamiltonian propagation.
\textit{HSR w/o Short-Impulse} is particularly harmful, showing that high-frequency stochastic excitation is crucial for handling noisy, bursty interaction patterns. 
\textit{HSR w/o Spectral Evolution} removing spectral evolution will lead to a moderate but consistent degradation on Amazon-Video-Games and MovieLens-1M, while the tiny gap on Amazon-Beauty indicates that for very short sequences the local and global components are partially redundant. \textit{HSR w/o Phase-Space Extrapolation} by disabling phase-space extrapolation and scoring only with the terminal position $\mathbf{q}_L$ yields a small but systematic drop, especially on Amazon-Video-Games, suggesting that the extrapolated state \cref{eq:extrapolation} indeed carries useful momentum information. Finally, \textit{HSR w/o Fourier Spectral} replacing the learned spectral kernel with the purely analytic Fourier Green’s function, although produces performance very close to HSR-Full, the full model still attains the best performance, indicating that the physically motivated denominator already provides a strong bias while the learnable numerator offers a consistent, if modest, refinement.

\subsection{Hyperparameter Sensitivity Analysis}
\subsubsection{Impact of Hamiltonian Spectral Block Number $(L)$}
We conduct experiment with varied number of Hamiltonian spectral blocks $(L)$ to examine the trade-off between network depth-accuracy for HSR. As shown in~\Cref{fig: hyper}(a) and (b), performance generally improves with additional blocks but saturates beyond the optimal levels, indicating diminishing returns for deeper architectures. Specifically, HSR achieves optimal performance with three Hamiltonian blocks for Beauty dataset, and four blocks for Game dataset.
The observed trend suggests that three or four blocks provide sufficient capacity to capture the complex temporal dynamics in user sequences while avoiding over-smoothing of user preference signals. Interestingly, performance slightly degrades with more than four blocks. The diminishing returns align with the physical intuition that  multiple Hamiltonian transformations can refine system dynamics, excessive transformations may lead to energy dissipation or phase-space distortion.

\subsubsection{Impact of Hidden Dimension $(d)$}
\label{subsec:hidden_dim_analysis}
We investigate the impact of hidden dimension size $(d)$ on model performance, a critical hyperparameter controlling model capacity. As shown in~\Cref{fig: hyper}(c) and (d), increasing $d$ from 64 to 256 yields substantial improvements: 11.5\% Hit@10 on Amazon-Beauty and 13.0\% Hit@10 on Amazon-Video-Games. However, further expansion to $d=512$ shows \textit{diminishing returns}, with performance gains dropping to 8.9\% and 10.2\% respectively.
This pattern aligns with the theoretical expectation: larger dimensions enhance representation capacity but increase risk of overfitting. 

\section{Discussions}
\label{subsec:ablation_layers}
\subsection{Qualitative Evaluation}
\begin{figure}[t]
  \centering
  \setlength{\abovecaptionskip}{0.2cm}
  \includegraphics[width=\linewidth]{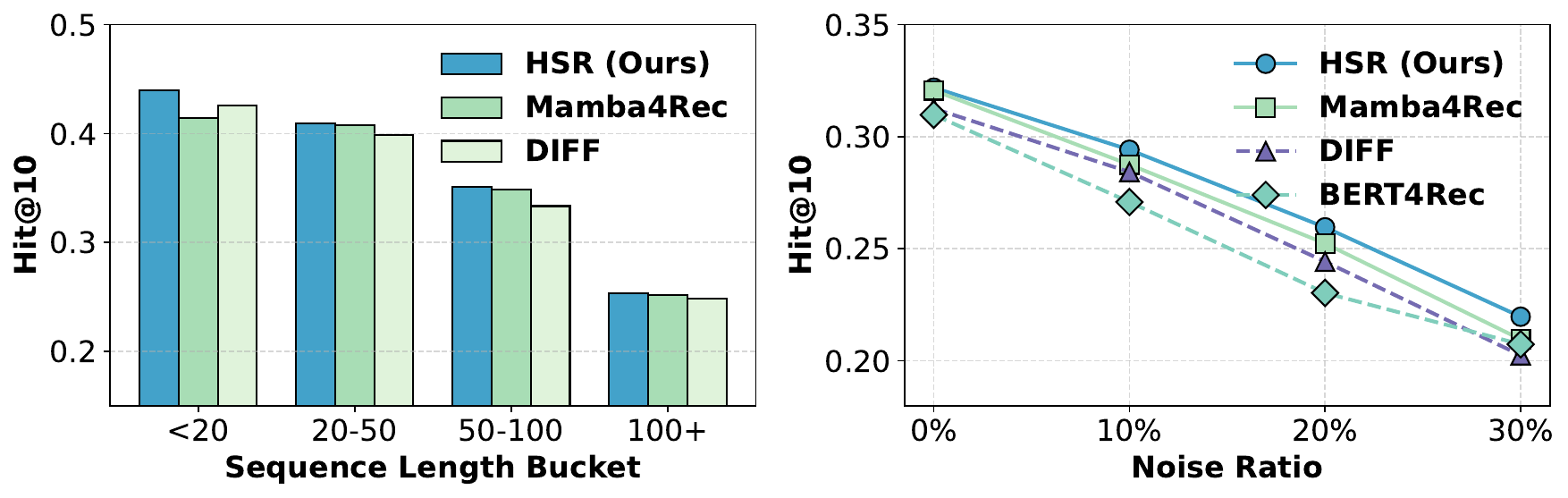}
  \caption{(a) Performance across Sequence Lengths and (b) Overall Robustness to Noise comparison.}
  \vspace{-5px}
  \label{fig: quant}
  \vspace{-8px}
\end{figure}

To further understand where the gains of HSR come from, we conduct two diagnostic studies whose results are summarized in Figure \ref{fig: quant}. 
\paragraph{Performance across Sequence Lengths.}
We bucket the MovieLens-1M test users by different interaction history length, and HSR consistently outperforms SOTA across all length ranges. On ultra-short sequences ($<$20), HSR still retains a clear margin, which we attribute to the local impulse refinement recovering short-term signal that spectral filtering alone would smooth out. On long sequences ($>$100), the gap against SSM and Transformer baselines is most pronounced: the Hamiltonian propagator encodes long-range temporal correlations through a frequency-dependent impedance $D_{n,i}^{(\ell)}$ rather than sequential state updates, avoiding the gradient decay and position-encoding saturation that limit first-order models on long histories.
\paragraph{Robustness to Interaction Noise.}
We inject synthetic noise by replacing a fraction ($0\%$--$30\%$) of each user's interactions with uniformly sampled items, simulating accidental clicks and exposure-driven noise. Figure~\ref{fig: quant}(b) shows that all baselines degrade substantially as noise ratio increases, while HSR exhibits the flattest degradation curve. This robustness has a direct physical interpretation: the damping coefficient $c_i$ in the spectral propagator attenuates high-frequency components of the driving force, which is precisely where noise injections concentrate in the frequency spectrum.


\begin{figure}[t]
  \centering
  \setlength{\abovecaptionskip}{0.2cm}
  \includegraphics[width=\linewidth]{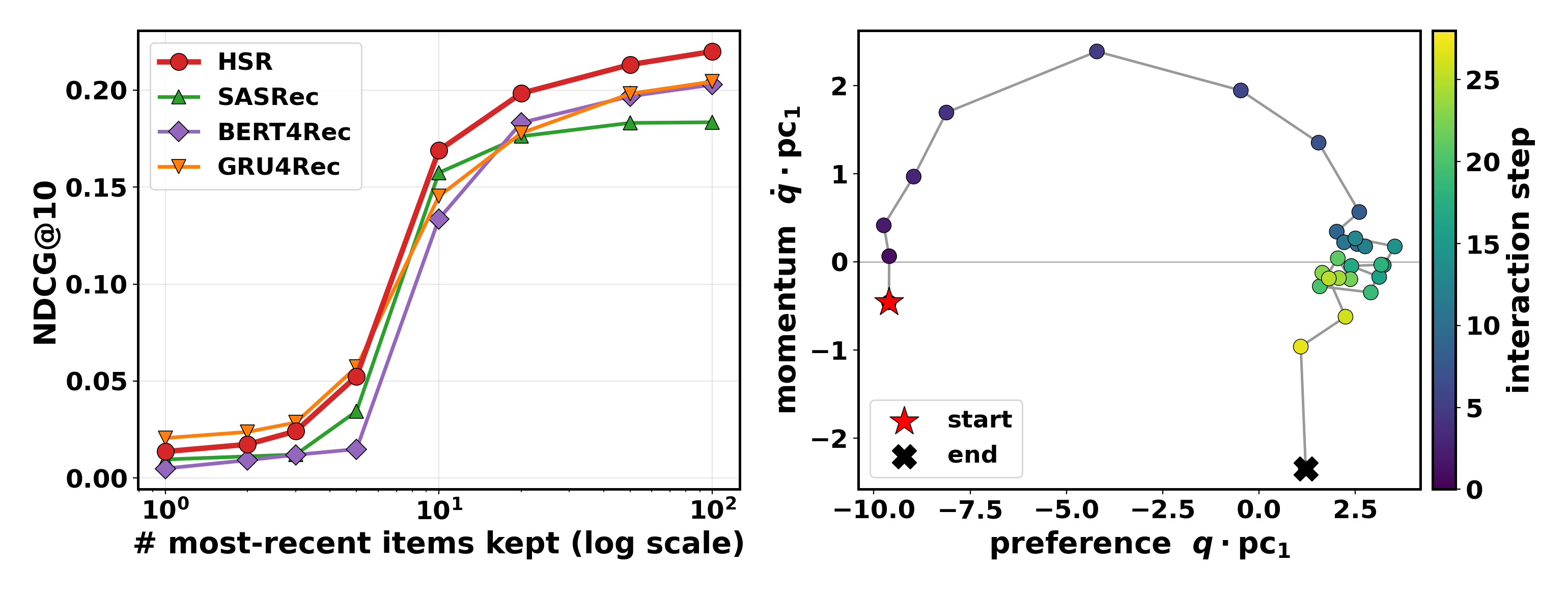}
  \caption{(a) Controlled cold-start on MovieLens-1M: each user's history is truncated to its $k$ most-recent items. (b) Phase portrait of user's latent trajectory: stable preference and momentum, colour-coded by step, $\star$ start and $\mathbf{\times}$ terminal state.}
  \label{fig:casestudy}
  \vspace{-8px}
\end{figure}

\paragraph{Cold-Start Robustness}
Figure~\ref{fig:casestudy}(a) reports NDCG@10 as each MovieLens-1M user's interaction history is truncated to its $k$ most recent items. HSR consistenly outerperforms the competing methods across nearly the entire range of history lengths, 
and its margin is largest in the cold-start regime, where only a few interactions are available. The local impulse refinement branch recovers short-term behavioral signals that may otherwise be smoothed out by global spectral propagation, 
and the momentum-based phase-space extrapolation provides 
a directional prior when context is scarce. As $k$ increases, 
the baselines partially catch up as they gain access to richer interaction histories; nevertheless, 
HSR remains best at most history lengths $k$, confirming that its second-order formulation is especially advantageous when the observed user history is short.

\paragraph{Interest Evolution in Phase Space}
Figure~\ref{fig:casestudy}(b) 
visualizes a representative user trajectory in HSR's latent phase space by plotting 
the stable-preference coordinate against its momentum along the principal 
preference axis. From the initial state ($\star$) the trajectory executes a large excursion 
associated with high momentum and then spirals inward, 
gradually converging to a compact attractor as the learnable damping dissipates energy. This orbit-and-settle behavior 
is the empirical realization of the second-order
dynamics modeled 
in HSR. In contrast, a conventional first-order model typically evolves through monotonic state transitions and cannot naturally represent the coupled preference-position and preference-velocity dynamics observed in the phase portrait. 

\paragraph{Comparison with Multi-Armed Bandit Recommender}
Earlier online multi-armed bandit recommenders, such as BMAB \cite{bmab}, 
model persistent and transient interests through a 
two-state bandit mechanism. To provide a direct comparison, Table~\ref{bmab} reports the average online next-item reward obtained over the top-$K$ most popular arms. HSR consistently outperforms BMAB across all values of $K$. This improvement can be attributed to the distinct 
modeling capacities of the two approaches. BMAB adopts a global, context-independent behavioral model that distinguishes persistent and transient interests primarily at the population level. 
In contrast, HSR demonstrates substantially stronger decision quality. The key advantage of HSR lies in its user-specific second-order latent state, which dynamically infers whether an individual user is currently exhibiting stable 
or exploratory behavior. This richer representation enables HSR to adapt recommendation decisions to user-specific preference dynamics, resulting in more accurate action selection and higher online reward.

\begin{table}[t]
  \centering
  \caption{Average reward $R(T)/N$ (top-1 among $K$ popular arms) on MovieLens-1M dataset.}
  \label{tab:bmab}
  \begin{tabular}{lccc}
    \toprule
    $K$ & 5 & 20 & 100 \\
    \midrule
    BMAB  & 0.0028 & 0.0112 & 0.0220 \\
    \textbf{HSR } & \textbf{0.0048} & \textbf{0.0154} & \textbf{0.0429} \\
    \bottomrule
  \end{tabular}
  \label{bmab}
\end{table}

\begin{figure}[t]
  \centering
  \setlength{\abovecaptionskip}{0.2cm}
  \includegraphics[width=\linewidth]{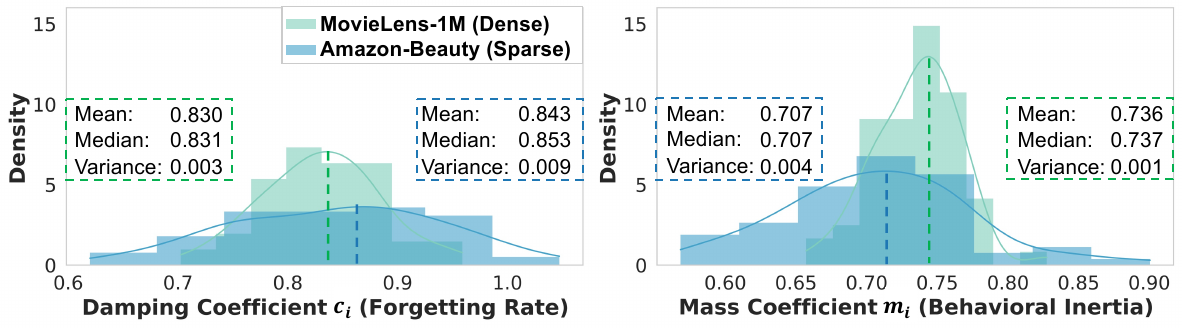}
  \caption{(a) Empirical Distribution of Learned Damping $\boldsymbol{c}_i$ and (b) Empirical Distribution of Learned Mass $\boldsymbol{m}_i$.}
  \label{fig: mass}
\end{figure}

\subsection {Interpretability of Learned Dynamical Parameters}

To examine whether the learned dynamical parameters align with intuitive user-behavior patterns, we analyze the distributions of mass and damping coefficients learned by HSR on different datasets.
Figure~\ref{fig: mass} shows the empirical distributions of the mass parameter $m_i$ and the damping parameter $c_i$ from the final HSR layer.

We observe that the Amazon-Beauty dataset, which characterized by extreme sparsity and short interaction histories, exhibit consistently higher damping values compared to MovieLens-1M.
This indicates faster attenuation of transient interests, which is consistent with highly volatile user behavior in sparse environments.
Conversely, MovieLens-1M exhibits larger mass values, implying stronger inertia and more persistent preference trajectories.
These trends suggest that HSR adapts its second-order dynamics to dataset characteristics in a behaviorally meaningful way, rather than learning arbitrary parameters.

\subsection{Comprehensive Efficiency Analysis}
\label{sec: cost}

\begin{table*}[htbp]
\centering
\caption{Comprehensive Efficiency Evaluation Based on Amazon-Beauty Dataset}
\label{tab: comprehensive}
\resizebox{\textwidth}{!}{%
\begin{tabular}{@{}lcccccc@{}}
\toprule
\textbf{Category} & \textbf{Metric} & \textbf{Mamba4Rec} & \textbf{DIFF} & \textbf{HSR} (Ours) & \textbf{Gain} & \textbf{ Improvement \%} \\
\midrule
\multirow{3}{*}{\textbf{Model Efficiency}} 
& Parameters (M) ↓ & 6.07 & 6.57 & \textbf{4.74} & \textbf{-1.33M vs Mamba4Rec} & \textbf{21.91\%} \\
& Model Size (MB) ↓ & 23.14 & 25.07 & \textbf{18.07} & \textbf{-5.07MB vs Mamba4Rec} & \textbf{21.91\%} \\
& Params/Throughput (Params/KQPS) ↓ & 178.8 & 367.2 & \textbf{74.7} & \textbf{-104.1 vs Mamba4Rec} & \textbf{58.22\%} \\
\midrule
\multirow{3}{*}{\textbf{Training Efficiency}}
& Time per Epoch (s) ↓ & 13.3 & 30.3 & \textbf{6.8} & \textbf{-6.5s vs Mamba4Rec} & \textbf{48.87\%} \\
& Eval Time per Epoch (s) ↓ & 0.8 & 1.0 & \textbf{0.6} & \textbf{-0.2s vs Mamba4Rec} & \textbf{25.00\%} \\
& Peak GPU Memory (GB) ↓ & 3.33 & \textbf{1.74} & 3.71 & \textbf{+1.97GB vs DIFF} & -113.22\% \\
\midrule
\multirow{3}{*}{\textbf{Inference Performance}}
& Single-user Latency (ms) ↓ & 0.025 & 0.054 & \textbf{0.014} & \textbf{-0.011
ms vs Mamba4Rec} & \textbf{44.00\%} \\
& Throughput (QPS) ↑ & 33,945 & 17,906 & \textbf{60,113} & \textbf{+26,168 vs Mamba4Rec} & \textbf{77.09\%} \\
& Full Sort Latency (ms) ↓ & 0.025 & 0.055 & \textbf{0.014} & \textbf{-0.011ms vs Mamba4Rec} & \textbf{44.00\%} \\
\midrule
\multirow{2}{*}{\textbf{Resource Utilization}}
& System Memory (GB) ↓ & 1.47 & 2.01 & \textbf{1.37} & \textbf{-0.1GB vs Mamba4Rec} & \textbf{6.80\%} \\
& CPU Utilization (\%) ↓ & \textbf{23.50} & 47.10 & 33.00 & \textbf{+9.5\% vs Mamba4Rec} & -40.43\% \\
\midrule
\multirow{2}{*}{\textbf{Efficiency Scores}}
& Latency Efficiency ↑ & 6.58 & 2.83 & \textbf{17.13} & \textbf{+10.55 vs Mamba4Rec} & \textbf{160.33\%} \\
& Throughput Efficiency ↑ & 5.59 & 2.73 & \textbf{13.39} & \textbf{+7.80 vs Mamba4Rec} & \textbf{139.53\%} \\
\bottomrule
\end{tabular}%
}
\label{tab: efficiency}
\end{table*}

Beyond accuracy and robustness, practical recommender systems critically depend on computational efficiency. We provide a thorough analysis of HSR’s efficiency across several dimensions in \Cref{tab: efficiency}. We compare these characteristics with competitive baselines to assess HSR’s practicality for large-scale deployment.

\subsubsection{Model Efficiency and Parameter Reduction}
HSR achieves a substantial reduction in parameter count and model size, which directly correlates with a lower memory and improved computational efficiency. This improved parameter efficiency arises from the structured nature of Hamiltonian propagation: rather than stacking many deep layers to achieve long-range modeling (as in Transformers), HSR relies on physically motivated propagation rules in the frequency domain. These rules implicitly encode global temporal dependencies without requiring large hidden dimensions or multi-head attention mechanisms. This reduction is primarily attributed to the architectural innovations that eliminate redundant transformations and leverage more compact sequence representations. 

\subsubsection{Training Efficiency and Memory Utilization}
A Hamiltonian spectral block requires one FFT and one inverse FFT per latent dimension. Each of these operations scales as $\mathcal{O}(T \log T)$, which is significantly more efficient than the $\mathcal{O}(T^2 \cdot D)$ complexity of the attention mechanism in Transformers. Although HSR introduces FFT-based computations, its overall training time remains competitive due to the shallow depth of the model and the efficiency of batched FFTs on modern GPUs. 
The training time per epoch is nearly halved compared to Mamba4Rec, while evaluation time is also reduced. This is largely due to optimized gradient flow and reduced computational graphs, which accelerate backpropagation. Interestingly, the peak GPU memory usage is higher than DIFF, which may appear counterintuitive. However, this can be explained by the model's use of intermediate representations that, while memory-intensive, enable faster convergence and better accuracy. 

\subsubsection{Inference Speed and Latency Improvements}
At inference time, HSR uses a forward propagation of the Hamiltonian spectral stack and one Euler extrapolation step. This makes inference efficient and stable, with runtime complexity dominated by the pair of FFT and iFFT operations.
A notable observation is the significant reduction in single-user latency and the corresponding increase in throughput. These gains stem from the model's ability to process user sequences more efficiently by minimizing sequential dependencies and enabling higher parallelism. 
Unlike Transformers that require full-sequence reprocessing for each inference step, HSR's structure allows cached intermediate states, low-latency spectral propagation, and parameter-light forward passes, which collectively support \textit{real-time inference} even under high request volume.

These results demonstrate that HSR offers a strong balance between expressiveness and computational efficiency. 
The efficiency scores, latency and throughput efficiency, further reinforce that the model achieves a Pareto improvement: better performance with lower resource consumption. 
\textit{HSR thus satisfies two key criteria for real-world sequential recommendation: (1) it significantly enhances predictive accuracy through principled second-order modeling; and (2) it remains computationally scalable, memory-efficient, and deployment-ready.}

\section{Conclusion}

We introduced \textbf{HSR}, a sequential recommender that recasts preference evolution as a dissipative Hamiltonian system in a latent phase space of position and momentum, departing from the first-order recurrences underlying existing Transformer- and SSM-based models. Three design choices make this principled yet practical: a physics-derived spectral propagator that solves the governing equation in $O(T\log T)$, a local impulse branch for short-term fluctuations, and a one-step phase-space extrapolation that turns the recovered momentum into the native mechanism for next-item prediction. Experiments on three benchmarks show consistent gains over state-of-the-art baselines, with damping acting as a physics-grounded low-pass filter against noise and the learned physical parameters adapting to dataset characteristics in behaviorally meaningful ways. Future work includes richer Hamiltonian parameterizations such as nonlinear potentials or adaptive dissipation, and extensions to multi-modal, context-aware recommendation.

\begin{acks}
The work described in this paper was supported, in part, by the Innovation and Technology Fund (Project: ITP/004/24TP) and by The Hong Kong University of Science and Technology (Grant: R9973). 
\end{acks}

\bibliographystyle{ACM-Reference-Format}
\bibliography{base}

\appendix

\end{document}